\documentclass{article}
\usepackage{techmech}
\usepackage{graphicx}
\begin{document}
\title{Phase Field Crystal Study of Symmetric Tilt Grain Boundaries of Iron}
\author{A. Jaatinen$^{1,2}$, C. V. Achim$^3$, K. R. Elder$^4$, T. Ala-Nissila$^{1,5}$ }
\maketitle
\begin{abstract}
We apply the phase field crystal model to study the structure and
energy of symmetric tilt grain boundaries of bcc iron in 3D. The
parameters for the model are obtained by using a recently developed
eight-order fitting scheme [A. Jaatinen \emph{et al.}, Phys. Rev. B
\textbf{80}, 031602 (2009)]. The grain boundary free energies we
obtain from the model are in good agreement with previous results
from molecular dynamics simulations and experiments.
\end{abstract}

\section{Introduction}

One of the challenges in modern computational materials science is
being able to access phenomena that take place on different time and
length scales. Recently, a new model called the phase field crystal
(PFC) model has been constructed that describes phenomena taking
place on atomic length scale but diffusive time scales
\citep{elder:2002}, the combination of which has been unaccessible
for molecular dynamics (MD) simulations using present day computers.
The PFC method is able to achieve this combination of scales by
replacing the individual atoms in MD simulations by a continuous
field that exhibits the periodic nature of the underlying atomic
lattice in the solid phase, but evolves in diffusive time scales. As
the original formulation of the PFC model relies on a
phenomenologically constructed free energy functional, it has proven
challenging to quantitatively relate the parameters that enter the
PFC model to the properties of real materials. In a recent study, we
have shown how the PFC model can be modified in such a way that many
relevant material properties can be included in the PFC free energy,
while still conserving most of the computational simplicity of the
model \citep{jaatinen:2009}. This can be achieved using the
so-called Eighth Order fitting (EOF) scheme. In the present work, we
use the EOF-PFC model to study the energy of symmetrically tilted
grain boundaries of bcc iron near its melting point. Results of our
calculations will provide an independent theoretical prediction for
the grain boundary energy of iron near its melting point. Perhaps
more importantly, they also provide indication on whether the EOF
scheme is a good candidate for future studies of grain boundary
related phenomena, such as grain boundary premelting
\citep{berry:2008,mellenthin:2008} and nanocrystalline deformation
\citep{elder:2004,stefanovic:2009}, with the PFC model.

\section{Model}

The phase field crystal (PFC) model describes crystallization
phenomena and has been applied to many different problems in
materials science. Originally, the model was phenomenologically
postulated in the spirit of the Ginzburg-Landau theory by
\cite{elder:2002}. The model consists of an order parameter field
that is driven by dissipative, conserved dynamics to minimize a free
energy functional, whose ground state exhibits a periodic structure
commensurate with the crystal symmetry of interest. More
specifically, the deterministic equation of motion for the locally
conserved order parameter field $n({\bf r})$ in these models is
given by
\begin{equation}
\label{eq:PFCdynamics} \frac{\partial n}{\partial t} = M \nabla^2
\frac{\delta F}{\delta n},
\end{equation}
where $M$ is mobility and $F$ is free energy of the system as a
functional of the field $n(\vec{r},t)$. The most usual choice of a
free energy functional for PFC studies is the form derived by
\cite{swift:1977} for a study of convective instabilities. Here we
briefly describe the procedure suggested by \cite{jaatinen:2009} for
obtaining a free energy that is capable of describing body-centered
cubic materials from a fundamental basis.

Derivation of our free energy starts from the classical
density-functional theory (DFT) of freezing, first pioneered by
\cite{ramakrishnan:1979}. The simplest free energy used in DFT of
freezing studies can be obtained by expanding the excess (over ideal
gas) contribution to the free energy around a uniform reference
density $\rho_0$, leading to
\begin{equation}
\label{eq:DFTn}   \frac{\Delta { F}[n({\bf r}) ]}{k_BT\rho_0} =
 \int d{\bf r}\left[ (1+n({\bf r})) \ln (1+n({\bf r}))-n({\bf r}) \right]
- \frac{1}{2}\int d{\bf r} \int d{\bf r}' n({\bf r})C(|{\bf r}-{\bf
r}'|)n({\bf r}'),
\end{equation}
where $\Delta F$ is the free energy difference between a given state
and the reference, $k_BT$ is the thermal energy and the
order-parameter field $n$ is related to the ensemble-averaged
one-particle density $\rho({\bf r})$ through $n({\bf r})=(\rho({\bf
r})-\rho_0)/\rho_0$. The function $C$ entering Eq. (\ref{eq:DFTn})
is the two-body direct correlation function (DCF) of the reference
liquid, which is related to the total pair correlation function, and
thus structure factor of the liquid through the well-known
Ornstein-Zernike relation.

In principle it is possible to use Eq. (\ref{eq:DFTn}) for PFC
studies; however, in practice solutions of this equation are very
sharply peaked in space (around atomic lattice positions) which
restricts numerical calculations to very small systems. To overcome
this difficulty, we make two approximations. Firstly, the non-local
part of $F$ is simplified by expanding the DCF in $k$-space up to
eighth order as
\begin{equation}
\label{eq:gradexpansion2} C(k) \approx C(k_m) - \Gamma \left(
\frac{k_m^2 - k^2}{k_m^2} \right)^2 - E_B \left( \frac{k_m^2 -
k^2}{k_m^2} \right)^4,
\end{equation}
where the parameter $k_m$ is the position of the first maximum in the
original DCF, and $\Gamma$ and $E_B$ are chosen such that the
expansion reproduces the position, height and curvature of the first
maximum, and the $k = 0$ value of the original DCF. This is achieved
by choosing
\begin{equation}
\label{eq:Gamma} \Gamma = - \frac{k_m^2 C''(k_m)}{8},
\end{equation}
where primes denote derivatives with respect to $k$, and
\begin{equation}
\label{eq:EB} E_B = C(k_m) - C(0) - \Gamma.
\end{equation}

As shown by \cite{jaatinen:2009}, this eighth order expansion
provides an excellent fit from $k=0$ to the first peak in $C(k)$ in
the case of iron. However, for the density oscillations whose wave
vector corresponds to a $k$ larger than $k_m$, the expanded free
energy now results in a significantly larger free energy penalty,
thus reducing stability of the solid phase. In order to partly
correct this error, and to further simplify the mathematical form of
the free energy, the local part is expanded in a fourth order power
series as
\begin{equation}
\label{eq:logexpansion} (1+n)\ln (1+n)-n \approx \frac{1}{2}n^2 -
\frac{a}{6}n^3 + \frac{b}{12}n^4.
\end{equation}
The coefficient $1/2$ for the second order term in this series is
obtained from a Taylor series in order to conserve the linear
stability of the model. For the third and fourth order terms we have
included constants $a$ and $b$, which are chosen such that the
solid will stabilize in the model, with the correct amplitude of
density fluctuations corresponding to the first star of reciprocal
lattice vectors, $u_s$. This is approximately achieved by choosing
them as
\begin{equation}
\label{eq:nonlinears} a=\frac{3}{2 S(k_m)u_s};\hspace{1cm}
b=\frac{4}{30S(k_m)u_s^2},
\end{equation}
where $S(k_m)=(1-C(k_m))^{-1}$ is height of the first peak in the
structure factor of the reference liquid \citep{jaatinen:2009}. With
these approximations, the free energy finally becomes
\begin{equation}
\label{eq:Feof} \frac{\Delta { F}[n({\bf r}) ]}{k_BT\rho_0} =
 \int d{\bf r}\left[ \frac{n({\bf r})}{2} \left( 1-C(k_m) + \Gamma \left(
\frac{k_m^2 + \nabla^2}{k_m^2} \right)^2 + E_B \left( \frac{k_m^2 +
\nabla^2}{k_m^2} \right)^4 \right)n({\bf r}) - \frac{a}{6}n({\bf
r})^3 + \frac{b}{12}n({\bf r})^4 \right].
\end{equation}
In our previous study we fitted the parameters of the current model
with molecular dynamics simulation data from \cite{wu:2007},
complemented by experimental data from \cite{itami:1984} and
\cite{jimbo:1993}, for iron. The resulting model parameters are
$k_m=2.985$ \AA$^{-1}$, $C(k_m)=0.668$, $\Gamma = 11.583$ and
$E_B=38.085$ for the DCF part, and $a=0.6917$ and $b=0.08540$ for
the non-linear part. The reference temperature and density,
necessary for converting model results into SI units are $T=1772$ K
and $\rho_0=0.0801$ \AA$^{-3}$. We have also shown that with these
parameters, the model produces very reasonable predictions for bulk
modulus, solid-liquid coexistence gap and low index surface free
energies. \citep{jaatinen:2009} In the next section, we will show
that this model is also capable of describing grain boundaries in
iron.

\section{Grain boundaries}

Using the model and parameters described in the previous section, we
have studied the symmetrically tilted grain boundaries of iron.
Three choices of tilting axis, $\left< 100\right >$, $\left<
110\right
>$ and $\left< 111\right >$, were studied. The calculations were performed at a density of
$(\rho-\rho_0)/\rho_0=0.030$, which is just above the density of the
solid at coexistence. A density close to the solid-liquid
coexistence was chosen due to the fact that the model parameters
have been obtained to reproduce the selected properties of iron
close to its melting point \citep{jaatinen:2009}.

First, we have found the lattice constant $a$ of the bcc phase that
minimizes the free energy at the density in question. Then, a
computational box of dimensions $(L_x,L_y,L_z)$, and periodic boundary
conditions in all directions was initialized by using the one-mode
approximation,
\begin{equation}
\label{eq:onemode} n({\bf r})=n_0 + (1+n_0) 4 u  \left[ \cos(qx)
\cos(qy)
 +  \cos(qx) \cos(qz) +  \cos(qy) \cos(qz)
\right],
\end{equation}
where $q=2 \pi/a_{bcc}$,  $a_{bcc}$ being the lattice spacing, and
$u$ is the amplitude of density fluctuations. First, a rotation to
the one-mode approximation was applied such that the tilt axis was
parallel to the $z$ axis of the computational box. Then, in the
region $x=0\ldots L_x/2$, the one-mode approximation was rotated in
the $(x,y)$ plane by an angle $-\theta/2$, and in the region
$x=L_x/2\ldots L_x$ by an angle $\theta/2$, creating two
symmetrically tilted grain boundaries, located at $x=0$ and
$x=L_x/2$, with a tilting angle $\theta$. The tilting angles
$\theta$ and box dimensions $(L_x,L_y,L_z)$ were chosen such that
the periodic pattern of the lattice continues smoothly over the
periodic boundaries in the $y$ and $z$ directions, and an integer
number of atomic planes was placed between the two grain boundaries
in the system. The system size in the direction of the tilting axis,
$L_z$, was chosen to be the closest distance between two similar
atomic planes in the $z$-direction. In the perpendicular directions,
$L_x$ and $L_y$ were varied from case to case, such that for small
angles, the simulation box contained several thousands of maxima,
while for large angle calculations, only hundreds of maxima were
required, due to the smaller spacing between dislocations at the
boundary.

\begin{figure}
\begin{center}

\includegraphics[width=100mm]{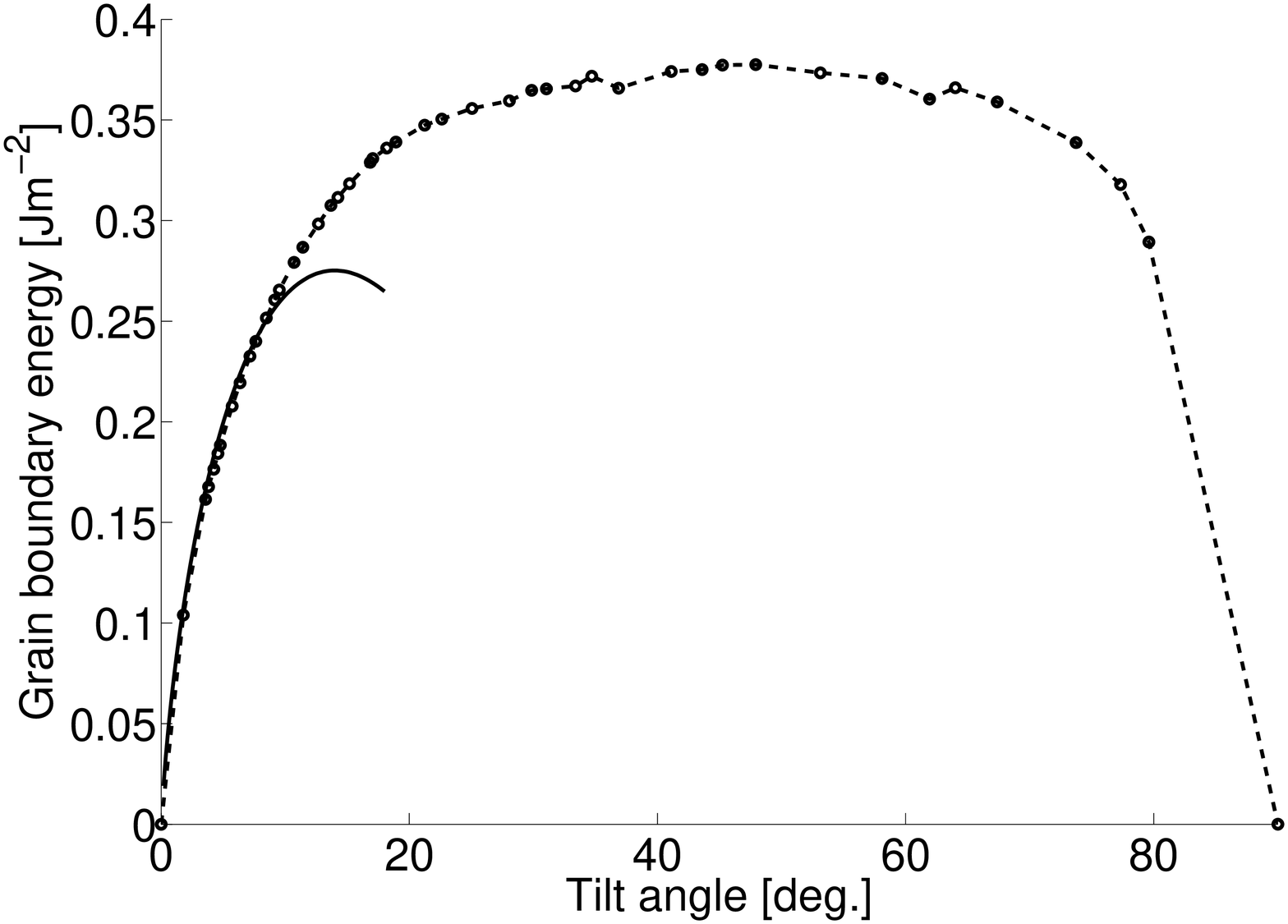}
\caption{Grain boundary free energy as a function of the
misorientation angle $\theta$ when the tilt axis is in the $\left<
100\right >$ direction. Solid line shows the best fit to the
Read-Shockley equation at small tilt angles. Dashed line is a guide
to the eye. \label{fig:GbeTheta100}}
\end{center}
\end{figure}

\begin{figure}
\begin{center}
\includegraphics[width=40mm]{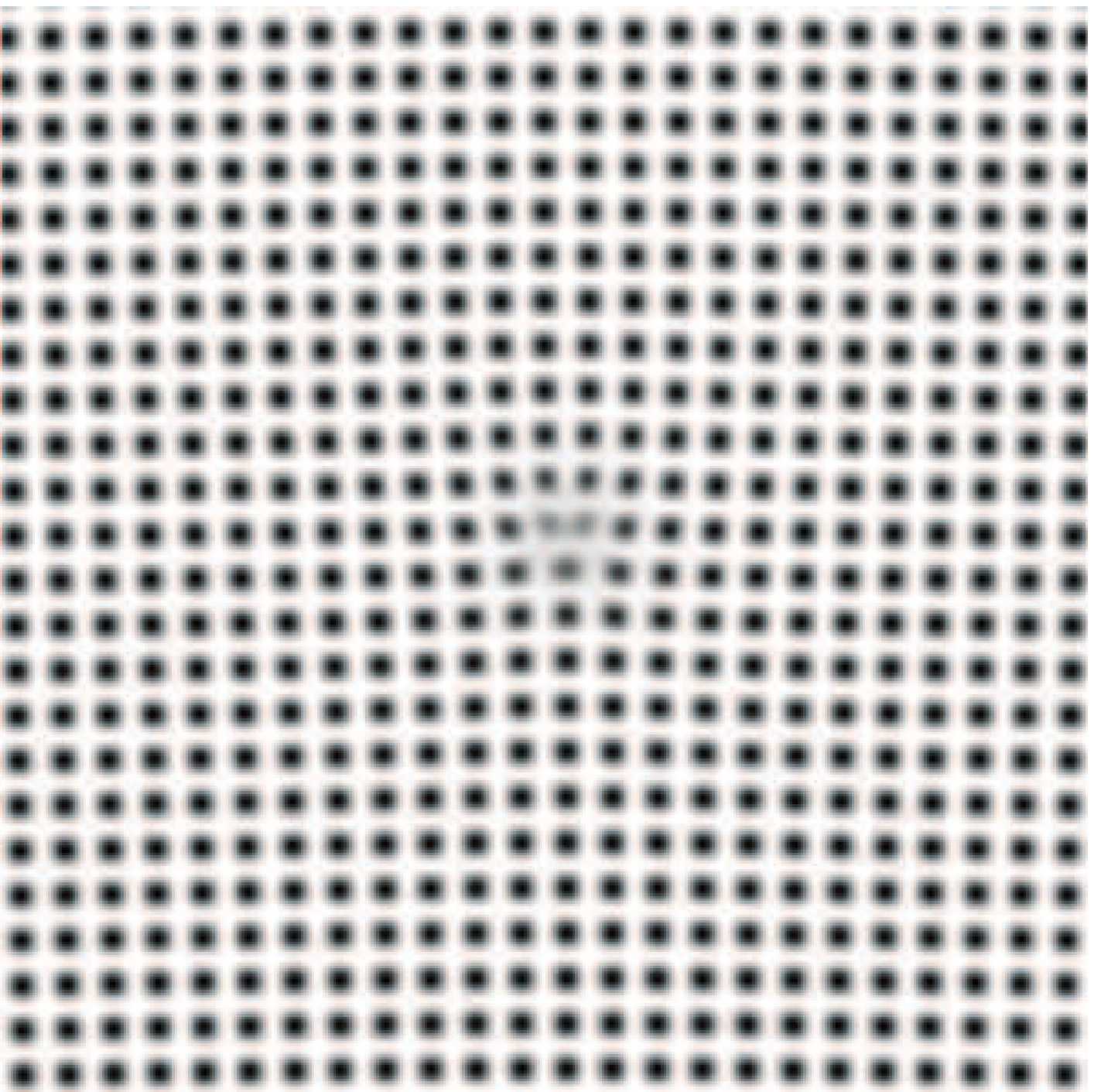}
\hspace{1cm}
\includegraphics[width=40mm]{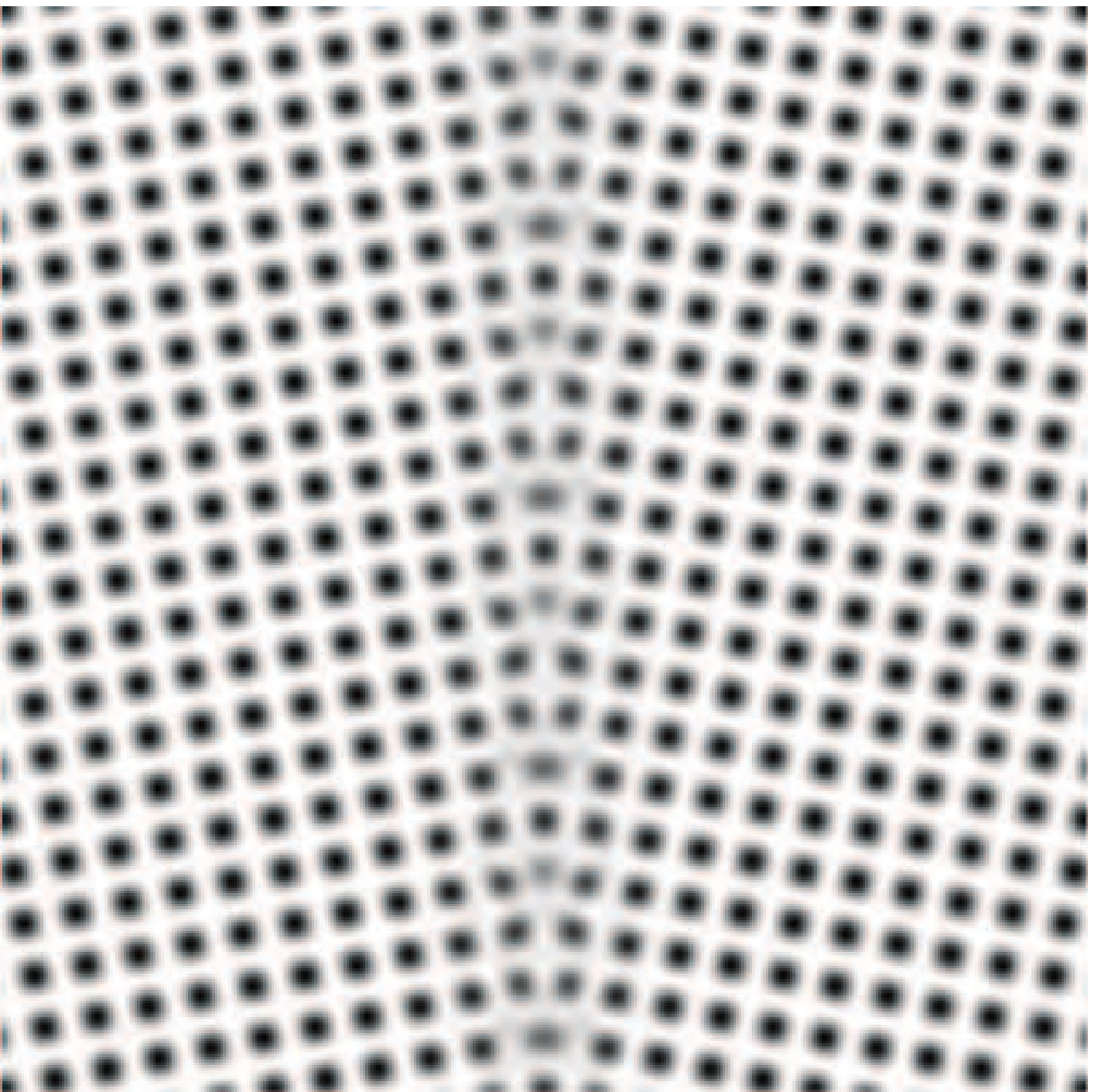}
\hspace{1cm}
\includegraphics[width=40mm]{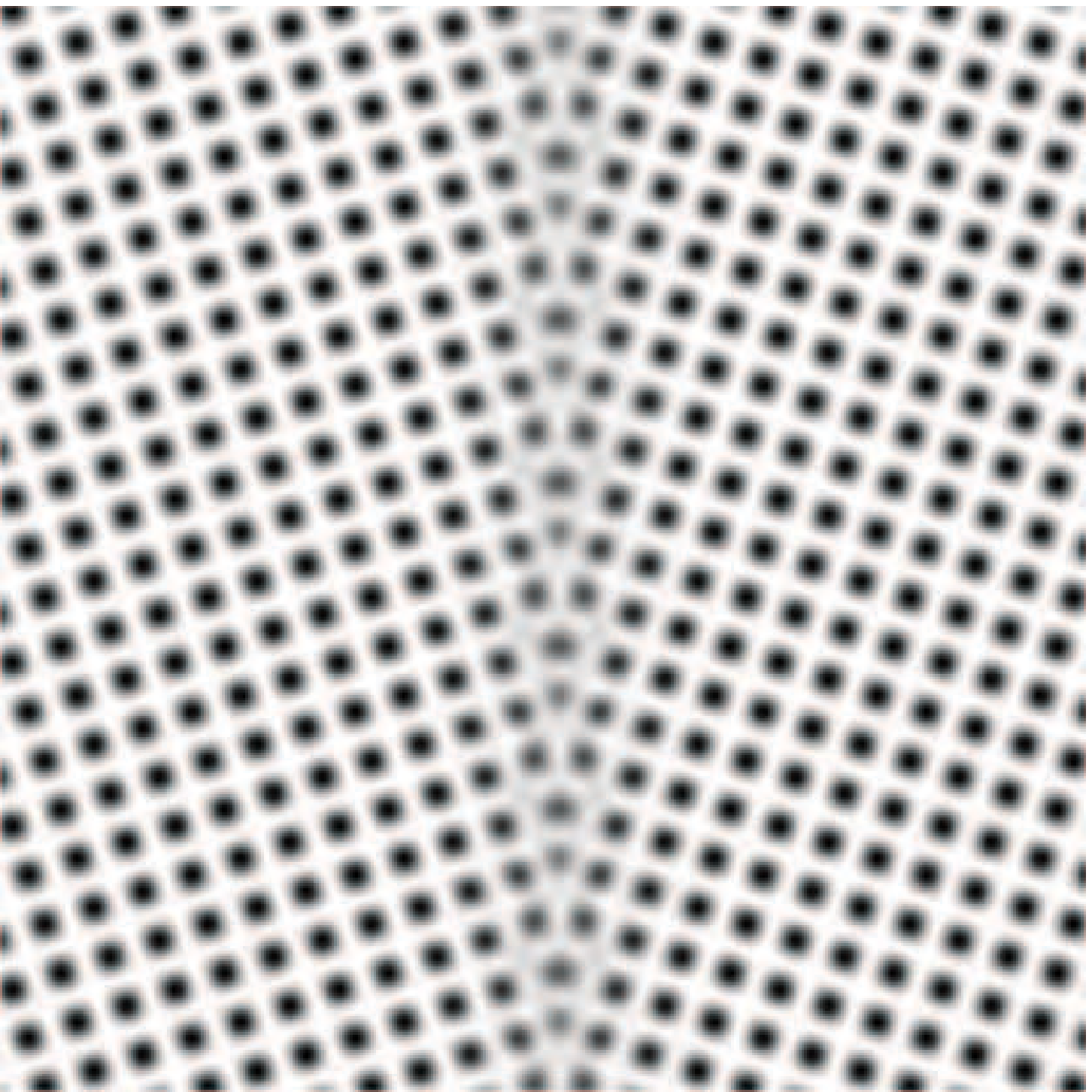}
\caption{Example final configurations from the $\left< 100\right
>$ case. The gray scale in these figures corresponds to $n(x,y,0)$.
On the left, we show the low-angle bcc$(0\ 64\
1)\left<100\right>\Sigma40975$ ($\theta=1.79 ^\circ$) boundary,
where one clearly sees the individual dislocation. The middle figure
shows a large angle bcc$(051)\left<100\right>\Sigma13$
($\theta=22.62 ^\circ$) boundary, and the figure on the right shows
the bcc$(031)\left<100\right>\Sigma5$ ($\theta=36.87 ^\circ$), for
which a small energy cusp was observed. \label{fig:Samples100}}
\end{center}
\end{figure}

In order to find the minimum energy of the system with the grain
boundaries, we have numerically integrated Eq.
(\ref{eq:PFCdynamics}) until convergence, using the semi-implicit
operator splitting method of \cite{tegze:2009}. The Laplace operator
is discretized in $k$-space as $\Delta_{\bf k}=-k^2$. Grain boundary
free energy per unit area, $E_{gb}$, is then obtained from
\begin{equation}
\label{eq:Egb} E_{gb}=L_z \frac{f-f_b}{2},
\end{equation}
where $f$ is the average free energy density of the final
configuration, and $f_b$ is free energy density of a non-rotated
system.

\begin{figure}
\begin{center}

\includegraphics[width=100mm]{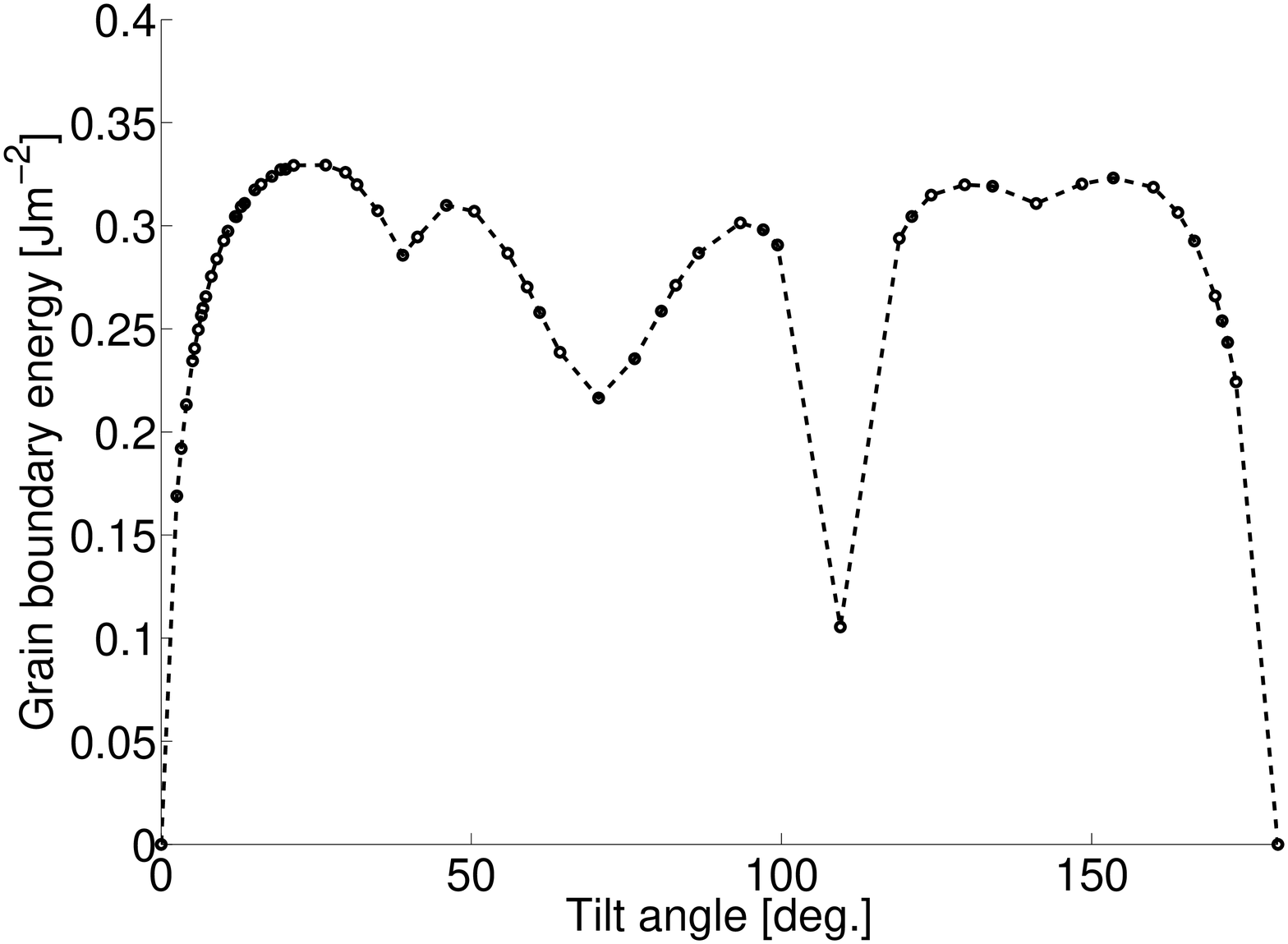}
\caption{Grain boundary free energy as a function of the
misorientation angle $\theta$ when the tilt axis is in the $\left<
110\right >$ direction. Dashed line is a guide to the eye.
\label{fig:GbeTheta110}}
\end{center}
\end{figure}

\begin{figure}
\begin{center}
\includegraphics[width=35mm]{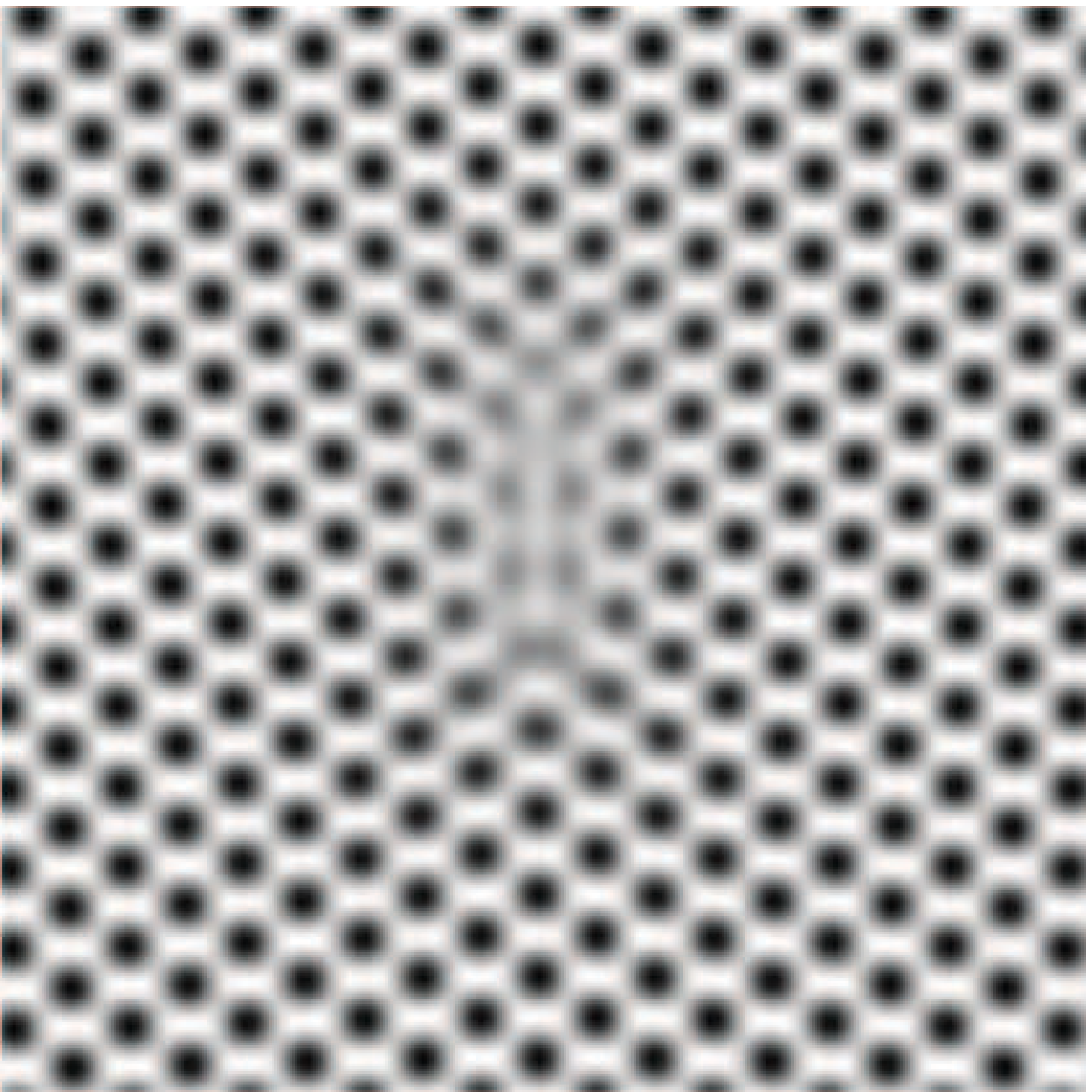}
\hspace{0.5cm}
\includegraphics[width=35mm]{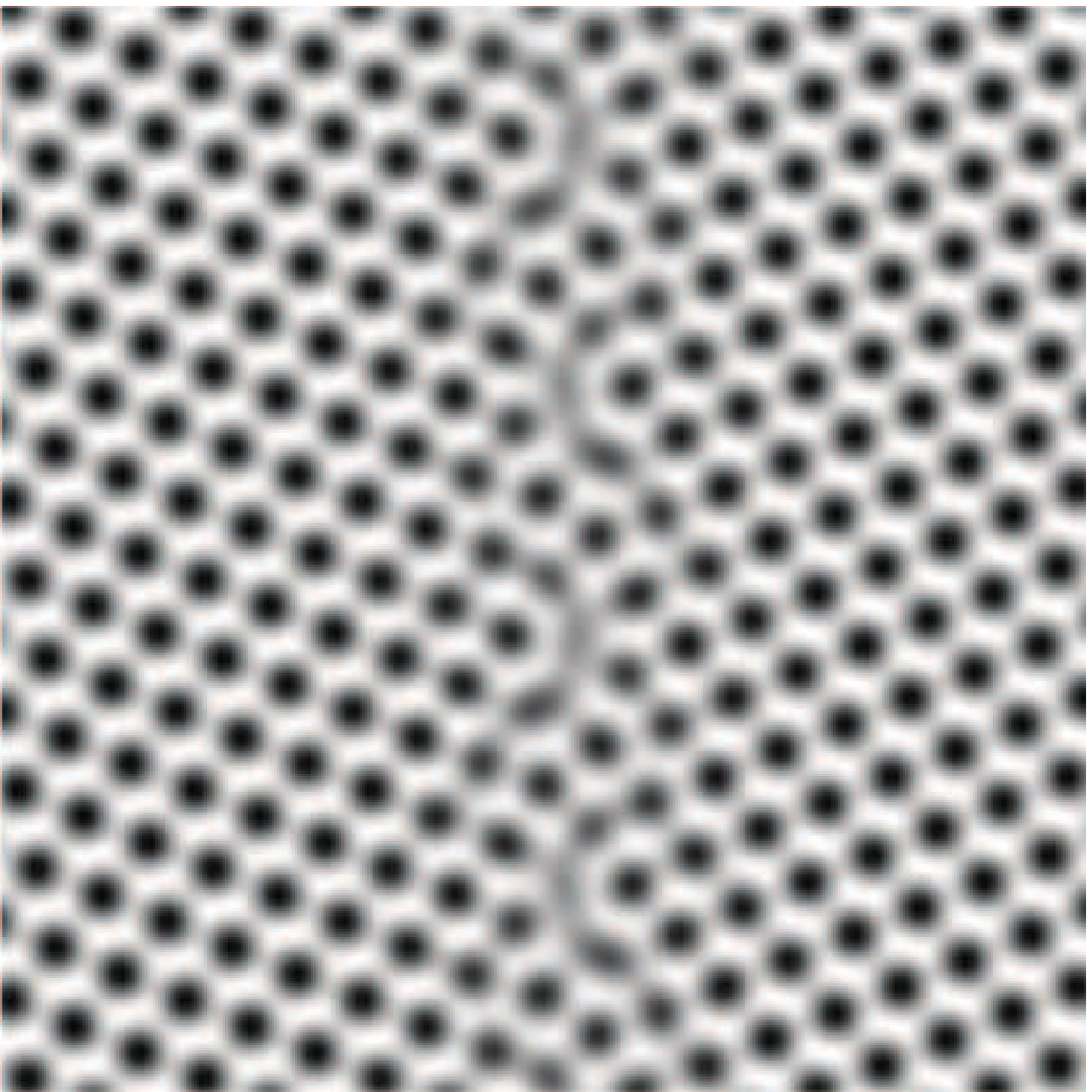}
\hspace{0.5cm}
\includegraphics[width=35mm]{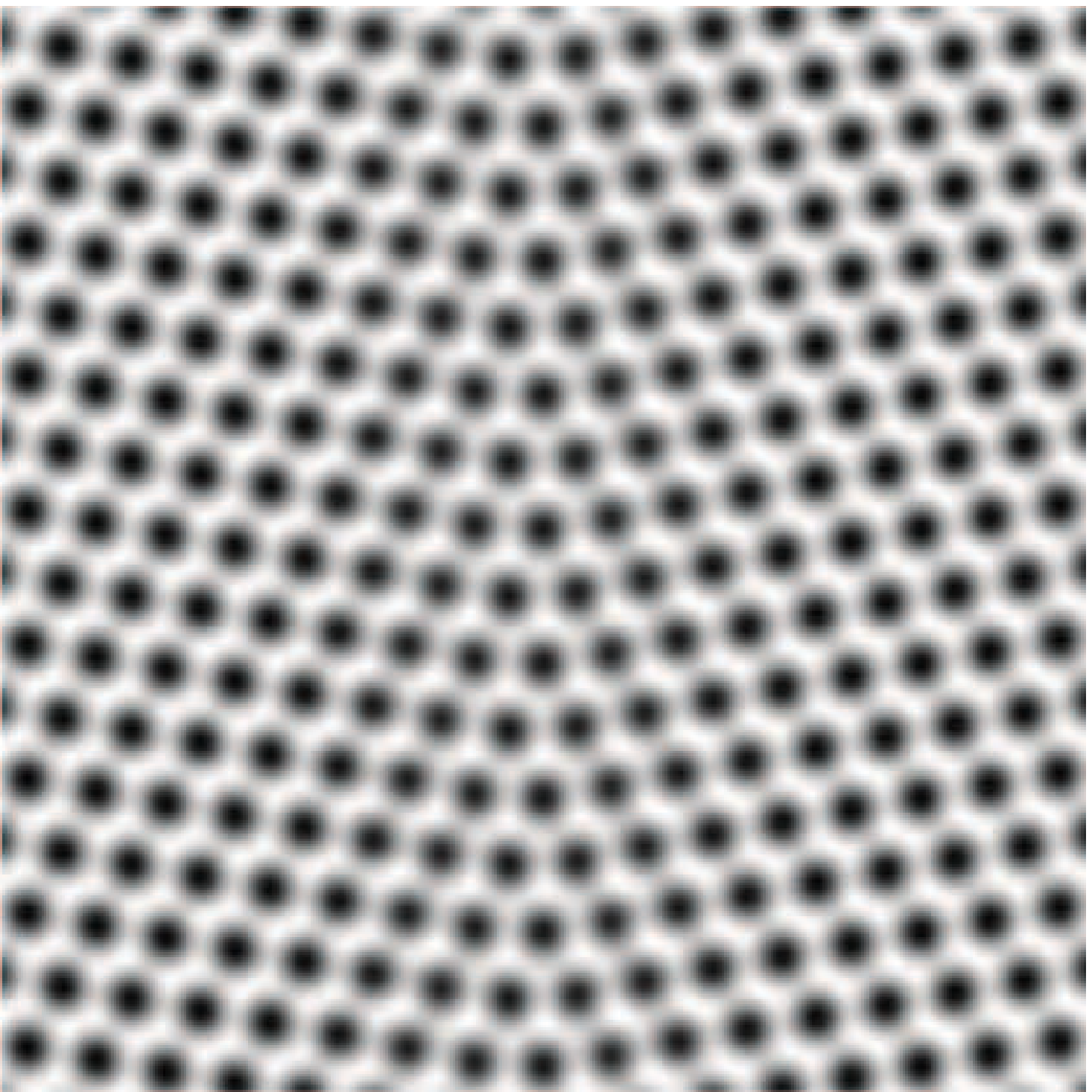}
\hspace{0.5cm}
\includegraphics[width=35mm]{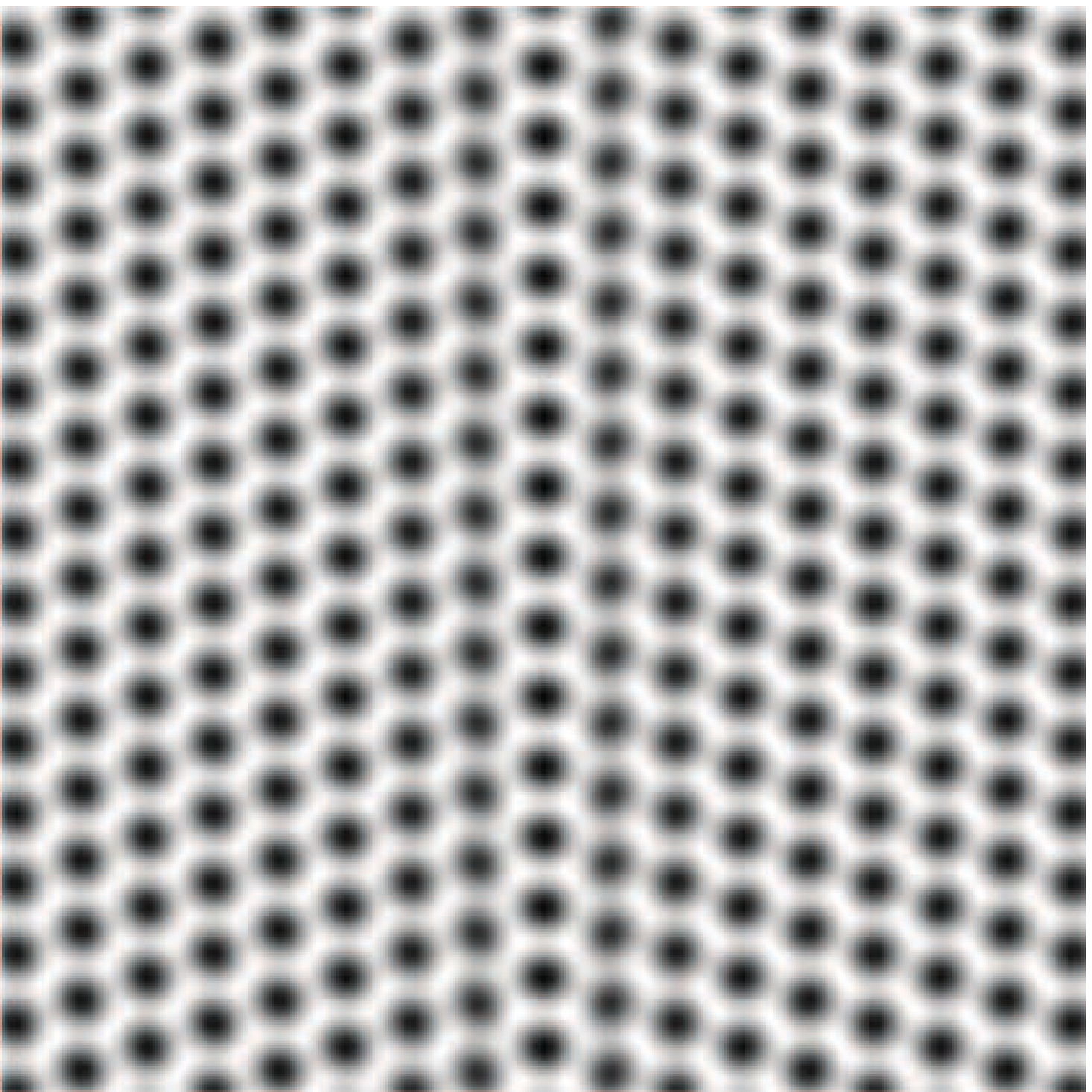}
\caption{Example final configurations from the $\left< 110\right
>$ case, with the gray scale corresponding to $n(x,y,0)$. From left
to right, the corresponding grain boundaries are bcc$(32\ 32\
1)\left<110\right>\Sigma2049$ ($\theta=2.53^\circ$),
bcc$(331)\left<110\right>\Sigma19$ ($\theta=26.52^\circ$),
bcc$(332)\left<110\right>\Sigma11$ ($\theta=50.47^\circ$) and
bcc$(112)\left<110\right>\Sigma3$ ($\theta=109.47^\circ$).
\label{fig:Samples110}}
\end{center}
\end{figure}

Fig. \ref{fig:GbeTheta100} shows the results of our grain boundary
free energy calculations, and Fig. \ref{fig:Samples100} shows three
example final configurations for the $\left< 100\right
>$ case. As in previous studies utilizing the original PFC free energy
functional \citep{elder:2002,elder:2004,mellenthin:2008}, we find
that for small angles the grain boundary energy closely follows the
expression by \cite{rs:50},
\begin{equation}
\label{eq:rs} E_{gb} = E_0 \theta \left(A-{\rm
ln}\left(\theta\right)\right),
\end{equation}
where 
$E_0$ is an energy scale that depends on elastic properties of the
material studied, and the constant $A$ depends on the dislocation
core energy. In addition, both of these constants depend on the
orientation of the grain boundary. In the present work, as we have
not developed numerical tools for quantifying all the properties
that enter the expressions for $E_0$ and $A$ \citep{rs:50}, we have
chosen to treat them as fitting parameters. The best fit to the
current data gives $E_0=1.1271$ Jm$^{-2}$ and $A=-0.41$.

For large values of $\theta$, the changes in $E_{gb}$ are moderate,
with the maximum value being 0.377 Jm$^{-2}$. By close inspection of
Fig. \ref{fig:GbeTheta100} one can distinguish small energy cusps at
angles corresponding to bcc$(031)\left<100\right>\Sigma5$
($\theta=36.87 ^\circ$) and bcc$(053)\left<100\right>\Sigma17$
($\theta=61.93 ^\circ$) grain boundary planes. Both of these energy
cusps correspond to a low coincidence site lattice (CSL) $\Sigma$.
However, for the other low-$\Sigma$ grain boundaries, for example
the bcc$(021)\left<100\right>\Sigma5$ ($\theta=53.13 ^\circ$), no
such cusps were observed. More generally, we do not observe any
clear correlation between the CSL $\Sigma$ and energies of large
angle grain boundaries. Thus, our results do not agree with
\cite{zhang:2005} who found, using molecular statics with modified
analytical embedded atom method, that the energy of symmetrically
tilted grain boundaries in iron increases with increasing $\Sigma$.
Our results agree much better with \cite{shibuta:2008} who have
performed molecular dynamics simulations with Finnis-Sinclair
potential to calculate the energies of different symmetric tilt
boundaries of iron in different temperatures. Their results are
qualitatively (but not quantitatively) very similar to ours, and
they also observe the small energy cusp at
bcc$(031)\left<100\right>\Sigma5$ present in our results as well.
However, unlike in our results, \cite{shibuta:2008} observe another
small energy cusp at bcc$(021)\left<100\right>\Sigma5$. The cusp we
observe at bcc$(053)\left<100\right>\Sigma17$ is not present in
their results.

\begin{figure}
\begin{center}
\includegraphics[width=100mm]{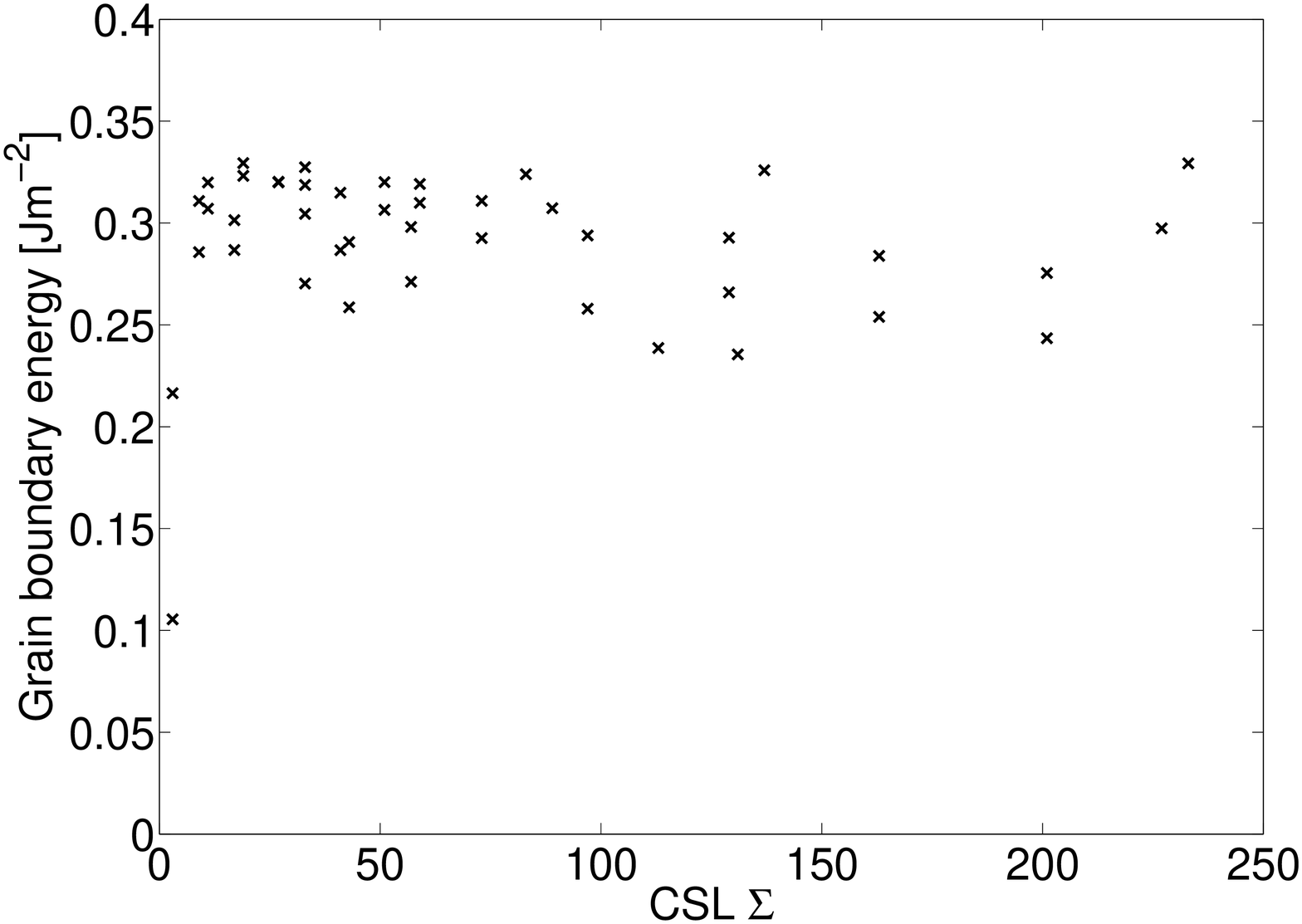}
\caption{Grain boundary free energy as a function of the coincidence
site lattice $\Sigma$ when the tilt axis is in the $\left< 110\right
>$ direction. \label{fig:110sigma}}
\end{center}
\end{figure}

Results of our grain boundary free energy calculations for the
$\left< 110\right >$ case are shown in Fig. \ref{fig:GbeTheta110},
and example final configurations for this case are shown in Fig.
\ref{fig:Samples110}. The low-angle behavior of $E_{gb}$ vs.
$\theta$ is qualitatively very similar to the $\left< 100\right
>$ case, but at large angles, interesting features are observed.
There is a deep minimum corresponding to
bcc$(112)\left<110\right>\Sigma3$ ($\theta=109.47^\circ$), which is
a twin boundary. The grain boundary energy in this minimum is only
approximately third of the maximum values observed. Other clear
local minima are observed at bcc$(221)\left<110\right>\Sigma9$
($\theta=38.94^\circ$), bcc$(111)\left<110\right>\Sigma3$
($\theta=70.53^\circ$) and bcc$(114)\left<110\right>\Sigma9$
($\theta=141.06^\circ$). Thus for $\left<110\right>$ grain
boundaries all angles with $\Sigma<10$ correspond to a local minimum
in grain boundary energy. However, as in the $\left< 100\right >$
case, we do not find a more general correlation between $\Sigma$ and
$E_{gb}$, as can be seen from Fig \ref{fig:110sigma}, where we plot
$E_{gb}$ vs. $\Sigma$ for the $\left< 110\right >$ case. In
comparison of our results with \cite{shibuta:2008}, we find that our
depth of the bcc$(112)\left<110\right>\Sigma3$ minimum is in an
excellent agreement with their work. However, their
bcc$(221)\left<110\right>\Sigma9$ minimum is much less pronounced
than in our work, and the remaining two of the local minima, that we
observed, were not observed in their study at all. Instead, they
observe a small energy cusp at bcc$(554)\left<110\right>\Sigma33$
($\theta=58.99^\circ$), which we did not find anomalous.

\begin{figure}
\begin{center}

\includegraphics[width=100mm]{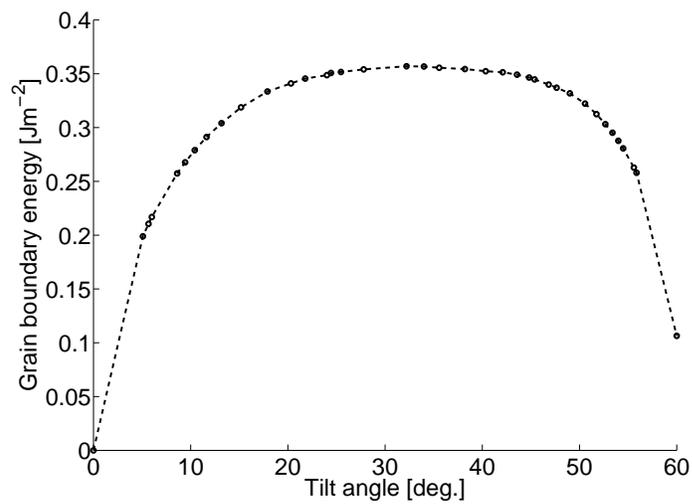}
\caption{Grain boundary free energy as a function of the
misorientation angle $\theta$ when the tilt axis is in the $\left<
111\right >$ direction. Dashed line is a guide to the eye.
\label{fig:GbeTheta111}}
\end{center}
\end{figure}

Results for the case where tilting axis is in the $\left< 111\right
>$ direction are shown in Fig. \ref{fig:GbeTheta111}. In this case,
the energy increases with increasing misorientation, without any
noteworthy energy cusps, until reaching a maximum value at $\theta =
32.20^\circ$. When further increasing $\theta$ from that maximum,
the grain boundary decreases, as the orientation gets closer to the
partial matching at $\theta=60^\circ$. Finally, at $\theta=60^\circ$
(i.e. bcc$(121)\left<111\right>\Sigma3$) the grain boundary energy
is 0.357 Jm$^{-2}$, which is approximately 30 \% of the maximum
value. The place of the maximum energy, magnitude of the
$\theta=60^\circ$ value compared to the maximum, and the absence of
energy cusps agree very well with the molecular dynamics results of
\cite{shibuta:2008}.

Experimentally, the large angle grain boundary energy of $\delta$
iron has been determined to be 0.468 Jm$^{-2}$ \citep{murr:1975}, in
good agreement with our maximum value. It is also worthwhile noting
that the agreement of our results with \cite{murr:1975} for the
ratio of large angle grain boundary energy to the solid-liquid
surface free energy is even better: $E_{gb}/\sigma_{s-l} \approx
2.16$ in the present model (solid-liquid surface energy from
\cite{jaatinen:2009}), where the values from \cite{murr:1975} give
$E_{gb}/\sigma_{s-l} \approx 2.29$. When comparing our values for
grain boundary energies with the results of the previously mentioned
atomistic calculations, we observe that the grain boundary energies
obtained by \cite{zhang:2005} are of the order of several Joules per
square meter, being significantly higher than the values reported
here or those reported by by \cite{shibuta:2008}.  It should be
noted that the calculations of \cite{zhang:2005} were conducted at
absolute zero temperature, not near the melting point, which perhaps
explains this discrepancy.  The grain boundary energies reported by
\cite{shibuta:2008}, on the other hand, reach a maximum of
approximately 1.6 Jm$^{-2}$ at moderate temperatures, rising to
approximately 2.4 Jm$^{-2}$ close to the melting point. A likely
reason for our grain boundary energies being closer to the
experimental values than those of \cite{shibuta:2008} is that we
are, in accordance to the experimentalists, actually measuring the
grain boundary \emph{free energy}, unlike \cite{shibuta:2008} who
only measure the excess energy of creating a grain boundary, without
taking into account the entropic part of the free energy.

\section{Conclusion}

We have applied the recently established EOF-PFC
model to study the structure and energy of symmetric tilt
grain boundaries in iron, with the axis of rotation parallel to
$\left<100 \right>$ and $\left<110 \right>$ directions. For small
misorientation angles, we found that the grain boundary energy
increases with increasing misorientation, in agreement with the
Read-Shockley equation. For large misorientation angles, we observed
local minima in the grain boundary energy vs. misorientation curve,
many of which have been observed in molecular dynamics simulations
as well. The local minima are associated with low values of the
coincidence site lattice $\Sigma$, but no general correlation
between $\Sigma$ and the grain boundary energy was found. In
general, the qualitative agreement between present results and
previous MD results at finite temperature was found
to be good. Quantitatively, the agreement of our result with an
experimental value for the large angle grain boundary energy is
better than that from the MD. As
our phase field crystal results agree well with previous MD
study of grain boundaries in iron, and an experimental
large angle grain boundary energy, they may even be considered an
independent prediction for the angular dependence of grain boundary
energy in iron. More importantly, they show the potential of the
EOF-PFC in modeling materials properties where grain
boundaries play an important role.

\section{Acknowledgements}
This work was supported in part by the Academy of Finland through
its Center of Excellence COMP grant, Tekes through its MASIT33
project, and joint funding under EU STREP Grant No. 016447 MagDot
and NSF DMR Grant No. 0502737. K.R.E. acknowledges support from NSF
under Grant No. DMR-0906676.

\bibliographystyle{techmech}
\bibliography{jaatinenetal.bbl}

\address{ $^1$Department of Applied Physics and
COMP Center of Excellence, Helsinki University of Technology, P.O. Box 1100, Helsinki FI-02015 TKK,
Finland\\ $^2$Department of Materials Science and Engineering,
Helsinki University of Technology, P.O. Box 6200, Helsinki FIN-02015
TKK,
Finland\\
$^3$Institut f\"ur Theoretische Physik II: Weiche Materie,
Heinrich-Heine-Universit\"at, Universit\"atsstrasse 1, D-40225
D\"usseldorf, Germany\\
$^4$Department of Physics, Oakland University, Rochester, Michigan
48309-4487, USA \\
$^5$Department of Physics, Brown University, Providence, Rhode
Island 02912-1843, USA\\
email: \texttt{Akusti.Jaatinen@tkk.fi} }

\end{document}